\documentstyle[aps,prb,twocolumn,epsf]{revtex} 

\begin{document}
\author{G. Gomila}
\address{Research Center for Bioelectronics and NanoBioScience, Departament\\
d'Electr\`{o}nica, Universitat de Barcelona, C/ Marti i Franques, 1, E-08028%
\\
Barcelona, Spain.}
\author{I. R. Cantalapiedra}
\address{Departament de F\'{\i}sica Aplicada, Universitat Polit\`{e}cnica de\\
Catalunya, Av. Dr. Mara\~{n}on, 44, E-08028 Barcelona, Spain.}
\author{T. Gonz\'{a}lez}
\address{Departamento de F\'{\i}sica Aplicada, Universidad de Salamanca, Plaza de la\\
Merced, s/n, E-37008 Salamanca,\ Spain.}
\author{L. Reggiani}
\address{INFM-National Nanotechnology Laboratory and Dipartimento di Ingegneria\\
dell'Innovazione, Universit\'{a} di Lecce, Via Arnesano s/n, I-73100 Lecce,\\
Italy.}
\title{Semiclassical theory of shot noise in ballistic $n^{+}-i-n^{+}$
semiconductor strucutres: relevance of Pauli and long range Coulomb
correlations.}
\date{\today}
\maketitle

\begin{abstract}
We work out a semiclassical theory of shot noise in ballistic $n^{+}-i-n^{+}$
semiconductor structures aiming at studying two fundamental physical
correlations coming from Pauli exclusion principle and long range Coulomb
interaction. The theory provides a unifying scheme which, in addition to the
current-voltage characteristics, describes the suppression of shot noise due
to Pauli and Coulomb correlations in the whole range of system parameters
and applied bias. The whole scenario is summarized by a phase diagram in the
plane of two dimensionless variables related to the sample length and
contact chemical potential. Here different regions of physical interest can
be identified where only Coulomb or only Pauli correlations are active, or
where both are present with different relevance. The predictions of the
theory are proven to be fully corroborated by Monte Carlo simulations.
\end{abstract}

\section{Introduction}

\label{Intro}

Ballistic conductors are characterized by an active region where carriers,
once injected by contacts, move without suffering any scattering from
contact to contact, i.e. the carrier mean free path is much longer than the
sample characteristic dimensions. In metals, ballistic transport is usually
studied in point contacts.\cite{Sharvin65} Such contacts allow the analysis
of elastic and inelastic scattering processes by means of the so called
point contact spectroscopy.\cite{Jansen80} Since the Fermi wavelength in
metals is very small ($\sim 0.5\ nm$), the nature of carrier transport is
semiclassical and quantum effects related to the wave nature of the
electrons can be disregarded. In semiconductors, ballistic transport has
been investigated in bulk materials,\cite{Heiblum85a,Levi85a} point contacts%
\cite{Trzcinski87} and two dimensional electron gases;\cite{Wees88} some
studies have allowed the development of ballistic emission spectroscopy\cite
{Kaiser88} for the analysis of semiconductor heterointerfaces. Moreover, a
variety of ballistic electron devices with promising performances have been
realized\cite{Heiblum85b,Levi85b} or proposed.\cite
{Natori94,Wernersson97,Gribnikov00} Since the Fermi wave length in
semiconductors can be as large as 40 $nm$, the nature of the carrier
transport can be either semiclassical or quantum depending on the
characteristic sample dimensions, $L$, with respect to the Fermi wave
length, $\lambda _{F}$. For $\lambda _{F}\ll L$ transport is semiclassical,
while for $\lambda _{F}\sim L$ transport is quantum.

From a fundamental point of view, the study of non-equilibrium electronic
noise (shot noise)\cite{Blanter00} of ballistic conductors in the
semiclassical transport regime offers a unique scenario where the
simultaneous effect of two fundamental physical interactions, namely the
long range Coulomb interaction and the Pauli exclusion principle, can be
investigated. On the one hand, Pauli correlations have an influence on the
carrier injecting statistics at the contacts. On the other hand, Coulomb
correlations can modify the carrier passing statistics inside the active
regions. The effect of both mechanisms results in values of the current
noise below the full poissonian value, i.e. in shot-noise suppression. Thus,
the non-equilibrium low frequency current spectral density is given by $%
S_{I}(0)=\gamma 2qI$, with $I$ being the current flowing trough the sample
and $\gamma $ a dimensionless parameter (Fano Factor) which takes values
below unity because of the negative correlations induced by both mechanisms.
The relative relevance of Coulomb and Pauli suppression in determining the
Fano factor depends on parameters like the sample length, the temperature,
the carrier density or the applied voltage, thus allowing the monitoring of
both correlation mechanisms.

The first theoretical analyses on the non-equilibrium noise properties of
ballistic conductors were performed by Van der Ziel and Bosman in $%
n^{+}-n^{-}-n^{+}$ ballistic semiconductor diodes\cite{vanderZiel82} and by
Kulik and Omel'yanchuk in ballistic metallic point contacts.\cite{Kulik83}
In the former case, by considering non-degenerate injection conditions, only
the effect of the Coulomb correlations was evidenced. In particular, by
using ideas borrowed from North's theory of vacuum tubes,\cite{North41} it
was shown that under space charge conditions current noise results
suppressed below the full Poissonian value (see also Refs. [%
\onlinecite{Gonzalez97,Bulashenko00a}]). In the later case, the high carrier
density of metals prevented the presence of significant space charge
effects, and hence the possibility to test the effect of the long range
Coulomb correlations. In this case, by means of a selfconsistent
semiclassical kinetic theory based on the Boltzmann-Langevin equation,\cite
{Kogan96} which included the effects of the Pauli principle, it was shown
that in the collision free regime ballistic metallic point contacts
displayed Nyquist noise with low frequency current spectral density given by 
$S_{I}(0)=4k_{B}T/R_{S}$, where $k_{B}$ is the Boltzmann constant, $T$ the
temperature and $R_{S}=mv_{F}/(q^{2}n)$ the Sharvin contact resistance, with 
$m$ being the electron effective mass, $v_{F}$ the Fermi velocity, $q$ the
unit charge, and $n$ the carrier concentration inside the sample. This
result is applied in the physical limit where $qV$, $k_{B}T\ll E_{F}$, and
arbitrary relation between $qV$ and $k_{B}T$, where $V$ is the applied bias
and $E_{F}$ the Fermi energy. These initial studies did not allowed the
study of the simultaneous effect of both Pauli and Coulomb correlation
mechanisms.

It was only very recently that ballistic structures including the
simultaneous effect of Pauli and Coulomb correlations have been analyzed.%
\cite{Naveh99,Gonzalez99,Bulashenko01} By making use of a semi-analytical
approach,\cite{Naveh99} Monte Carlo (MC) simulations,\cite{Gonzalez99} and
an asymptotic analytical approach,\cite{Bulashenko01} non-equilibrium noise
properties of ballistic semiconductor structures have been investigated. It
was shown that these structures, by allowing both space charge effects and
degenerate injection statistics, were unique in order to study the
simultaneous effect of both Pauli and Coulomb correlations.\cite{remarkFrac}
However, the existing studies, being only valid under some limit conditions
or performed for some specific sets of system parameters, do not offer a
complete overview of the relative relevance of Pauli and Coulomb
correlations in the whole range of system parameters and applied bias. This
lack of a general overview can prevent the designing of suitable
experimental strategies to test the theoretical predictions. The aim of the
present work is precisely to address this issue.

To this purpose we work out a general theory for the low frequency
shot-noise properties of ballistic $n^{+}-i-n^{+}$ semiconductor diodes that
consistently describes also current-voltage characteristics. We consider
ligthly doped bulk semiconductor materials since at low temperatures the
bulk electrons are trapped by their parent donors, neutralizing them and
thus minimizing both the $e-e$ interaction and ionized impurity scattering,
thus allowing long mean free paths up to relatively high electron energies.%
\cite{Brill96,Xie99} The theory presented accounts for both the Pauli
exclusion principle and the long range Coulomb interaction, and is applied
to the whole range of system parameters (sample length, temperature, contact
doping) and external bias. In particular, the theory allows us to study in a
unifying framework the transition from: (i) nondegenerate to degenerate
injection conditions, (ii) short to (asymptotically) long sample lengths
and, (iii) low to (asymptotically) high applied voltages. By overcoming the
limitations of the previous existing theories, the present study allows us
to compress into a general scheme the full scenario displayed by the Pauli
and Coulomb correlations in these ballistic structures. This scheme,
validated and tested with a wide set of MC simulations, is believed to be of
relevant assistance in designing future investigations on an experimental
and/or simulation basis.

The paper is organized as follows. In Sec.~\ref{System} we describe the
system under study. In Sec.~\ref{Model} we present the physical model used
to analyze the non-equilibrium noise properties. In Sec.~\ref{solution} the
analytical expressions obtained for the transport and noise properties are
reported, and their validity is checked by means of MC simulations. In Sec.~%
\ref{General} we propose a general scheme able to provide a systematic
physical picture of the shot-noise properties of two-terminal ballistic
conductors, and check its reliability by means of the developed theory.
Finally, in Sec.~\ref{Conclusions} we summarize the main results of the
paper. The appendix is devoted to technical derivations.

\section{System under study}

\label{System}

\begin{figure}[t]
\centerline{
\epsfxsize=10cm \epsffile{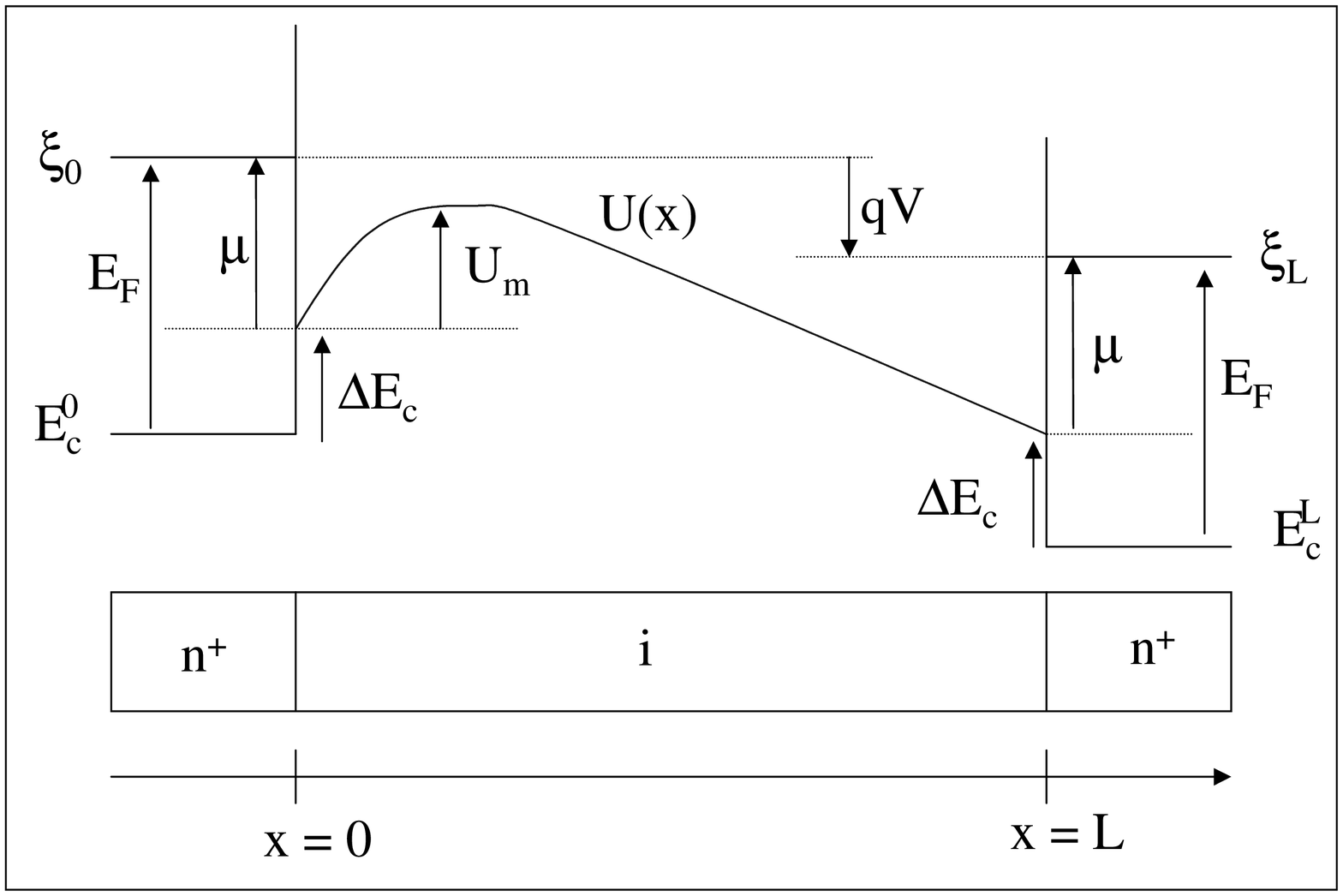}}
\caption{Schematic band-diagram of the ballistic structure under study.}
\label{FigSystem}
\end{figure}

The system under study is the $n^{+}-i-n^{+}$ ballistic semiconductor diode
showed in Fig.~\ref{FigSystem}. 

It consists of an undoped semiconductor of
length $L$, where injected carriers move in the total absence of scattering
events, sandwiched between two highly doped regions which act as ideal
injecting contacts as specified later. The contacts and the ballistic region
are taken of the same material (homodiode) or of different material
(heterodiode). In the former case the band offset $\Delta E_{c}\equiv
U(0)-E_{c}^{0}=$ $U(L)-E_{c}^{L}$ (a symmetric structure is assumed for
simplicity) vanishes, while in the later case it takes a finite value. The
ballistic region is taken to be perfectly coupled with the contacts, thus no
reflections take place at the interfaces. The contacts are assumed to be in
quasiequilibrium at the given temperature, and at electrochemical potentials 
$\xi _{0}$ and $\xi _{L}$, with $\xi _{L}-\xi _{0}=qV$. The voltage drop in
the contacts is assumed to be negligibly small (Ohmic contacts), and hence
all the band bending occurs in the active region of the sample. Under these
assumptions, the effective contact chemical potential $\mu $, defined as $%
\mu \equiv \xi _{0}-U(0)=\xi _{L}-U(L)$ is independent of the applied bias.
Note that for the case of an homodiode $\mu $ coincides with the contact
Fermi energy, $E_{F}\equiv \xi _{0}-E_{c}^{0}=\xi _{0}-E_{c}^{L}$, while for
the case of an heterodiode one has $\mu =E_{F}-\Delta E_{c}$. We note the
existence of a maximum with amplitude $U_{m}$ in the potential energy along
the active region due to the presence of space charge. The structure is
assumed to be sufficiently thick in the transversal directions so as to
allow for a 1D electrostatic treatment. Thus the system is 1D in real space
and 3D in momentum space.\cite{Bulashenko00a} The carriers are injected into
the active region in accordance with the equilibrium Fermi-Dirac
distribution function and satisfying the Pauli exclusion principle, thus
following the corresponding binomial injection statistics.\cite
{Levitov93,Gonzalez99b} For simplicity, a single spherical parabolic band is
assumed.

The parameters describing the model system introduced above are therefore
the length of the conductor, $L$, the cross sectional area $A$, the
effective contact chemical potential, $\mu $, the temperature, $T$, the
static dielectric constant of the medium, $\epsilon $, the electron
effective mass, $m$, and the charge carriers $q$. The independent-variable
parameter is the applied voltage, $V$.

As mentioned above, the correlations between different current pulses in the
present system 
is twofold. On the one hand, Pauli correlations have an influence on the
contact injecting statistics. Carriers above the Fermi energy, at the tail
of the Fermi distribution at the contacts, are injected obeying Poissonian
statistics due to the negligible influence of the Pauli principle at such
high energies. Carriers near and below the Fermi energy are injected
following a binomial (sub-Poissonian) statistics due to the increasing
influence of the Pauli principle at the highly occupied states. On the other
hand, Coulomb correlations, through the potential energy maximum $U_{m}$
associated with the presence of space charge, can modify the carrier passing
statistics inside the active region. Indeed, the fluctuations of $U_{m}$
originated by the passage of carriers over the maximum modulate the
transmission of further carriers (spacing them out more regularly in time)
and smooth out the current fluctuations imposed by the random injection at
the contacts. The relative importance of the two types of correlating
mechanisms depends on the values of $L$, $\mu $ and $V$.

\section{Physical model}

\label{Model}

Within a semiclassical approach the description of the transport and noise
properties of the system under study can be carried out by means of the
Vlasov equation, self-consistently coupled to the Poisson equation, and
supplemented by appropriate {\em fluctuating} boundary conditions.\cite
{Kogan96} In the one dimensional approximation followed here, the
Vlasov-Poisson system of equations reads (in dimensionless units, as
specified later): 
\begin{equation}
\left[ \frac{\partial }{\partial t}+v_{x}\frac{\partial }{\partial x}-\frac{%
\partial U(x,t)}{\partial x}\frac{\partial }{\partial v_{x}}\right]
F(x,v_{x},t)=0\text{ ,}  \label{Vlasov}
\end{equation}
\begin{eqnarray}
\frac{\partial ^{2}U(x,t)}{\partial x^{2}} &=&-n(x,t),  \label{Poisson} \\
n(x,t) &=&\int_{-\infty }^{+\infty }dv_{x}F(x,v_{x},t)\text{ ,}
\label{density}
\end{eqnarray}
where $t$ is the time variable, $x$ the spatial coordinate, $v_{x}$ the
velocity in the $x$ direction, $U(x,t)$ the potential energy, $F(x,v_{x},t)$
the distribution function integrated over transversal momentum directions
and $n(x,t)$ the carrier density. In the Poisson equation the contribution
of the intrinsic free carrier density is neglected in comparison to the
injected carriers. Moreover, the boundary conditions for the potential
energy read 
\begin{equation}
U(0,t)\equiv U_{0}=0\text{;\qquad }U(L,t)\equiv U_{L}=V\text{ ,}  \label{Ubc}
\end{equation}
where we have assumed that the applied bias is fixed by a low impedance
external circuit so that neither the electric potential nor the contact
potential energies fluctuates, i.e., $\delta U_{0}(t)=\delta U_{L}(t)=\delta
V(t)=0$. Accordingly, the boundary conditions for the distribution function
at the contacts read: 
\begin{eqnarray}
F(0,v_{x},t) &=&\overline{F}(v_{x})+\delta F_{0}(v_{x},t)\text{; \qquad }%
v_{x}>0\text{ ,}  \label{Fbc0} \\
F(L,v_{x},t) &=&\overline{F}(v_{x})+\delta F_{L}(v_{x},t)\text{; \qquad }%
v_{x}<0\text{ .}  \label{FbcL}
\end{eqnarray}
These boundary conditions consist of two contributions. The first is of
deterministic nature, and gives the average value of the distribution
function. The second is of stochastic nature, and describes the random
injection of carriers. The average distribution function is given by: 
\begin{eqnarray}
\overline{F}(v_{x}) &=&\int_{0}^{+\infty }d\varepsilon _{\perp }f_{FD}\left(
\varepsilon _{\perp }+\varepsilon _{x}-\mu \right)  \nonumber \\
&=&\ln \left( 1+e^{-\varepsilon _{x}+\mu }\right) \equiv f_{c}\left(
\varepsilon _{x}-\mu \right) \text{ ,}  \label{Fbcav}
\end{eqnarray}
with $\varepsilon _{x}=\frac{1}{2}v_{x}^{2}$ and $\varepsilon _{\perp }=%
\frac{1}{2}v_{\perp }^{2}$ being the longitudinal and transversal kinetic
energy, respectively, and where 
\begin{equation}
f_{FD}\left( \varepsilon \right) =\frac{1}{1+e^{\varepsilon }}  \label{FFD}
\end{equation}
is the Fermi-Dirac distribution function. Note that due to the integration
over the transverse momentum directions the effective one dimensional
contact distribution function is $f_{c}\left( \varepsilon \right) $ instead
of $f_{FD}\left( \varepsilon \right) $. Moreover, the fluctuating
contributions $\delta F_{0}(v_{x},t)$ and $\delta F_{L}(v_{x},t)$ have zero
mean and low frequency spectral density given by\cite{Bulashenko01} 
\begin{eqnarray}
&&2\int_{-\infty }^{+\infty }\overline{\delta F_{a}(v_{x},t)\delta
F_{a^{\prime }}(v_{x}^{^{\prime }},t^{\prime })}dt=  \nonumber \\
&&\frac{\partial f_{c}\left( \varepsilon _{x}-\mu \right) }{\partial \mu }%
\delta _{a,a^{\prime }}\delta (\varepsilon _{x}-\varepsilon _{x}^{\prime
})\delta (t-t^{\prime })=  \nonumber \\
&&f_{FD}\left( \varepsilon _{x}-\mu \right) \delta _{a,a^{\prime }}\delta
(\varepsilon _{x}-\varepsilon _{x}^{\prime })\delta (t-t^{\prime })\text{ .}
\label{Fccorr}
\end{eqnarray}
Equations (\ref{Vlasov})-(\ref{Fccorr}) constitute the complete set of
equations to study the noise properties of the ballistic structure described
in Sec.~\ref{System}

In the previous set of equations, and in what follows, we use dimensionless
variables to simplify the notation. The dimensionalizing factors used for
energy, length, carrier density, velocity, distribution function, electric
potential, electric field, electric current, resistance and current spectral
density are, respectively,

\begin{eqnarray}
U_{0} &=&k_{B}T\text{ ; }L_{0}=\sqrt{\frac{\epsilon k_{B}T}{q^{2}N_{0}}}%
\text{; }N_{0}=4\pi \frac{(mk_{B}T)^{3/2}}{h^{3}}\text{ ;}  \nonumber \\
\text{ }v_{0} &=&\left( \frac{k_{B}T}{m}\right) ^{1/2}\text{; }F_{0}=\frac{%
N_{0}}{v_{0}}\text{ ; }V_{0}=\frac{k_{B}T}{q}\text{ , }  \label{Dimen1} \\
E_{0} &=&\sqrt{\frac{N_{0}k_{B}T}{\epsilon }}\text{ ; }I_{0}=qAN_{0}v_{0}%
\text{ ; }R_{0}=\frac{V_{0}}{I_{0}}\text{ ; }S_{I_{0}}=2qI_{0}\text{,} 
\nonumber
\end{eqnarray}
where $h$ is the Planck constant. With these dimensionalizing factors the
dimensionless parameters describing the system are $L/L_{0}\rightarrow $ $L$
and $\mu /k_{B}T\rightarrow \mu $, and the dimensionless independent
variable is $qV/k_{B}T\rightarrow V$. The physical meaning of the different
parameters is the following: $N_{0}$ is the effective density of states in
the conduction band, $L_{0}$ is the Debye screening length associated to $%
N_{0}$, $U_{0}$ is the thermal energy, $v_{0}$ is the thermal velocity, $%
V_{0}$ is the thermal voltage, $E_{0}$ the thermal electric field, $I_{0}$
the current associated to $N_{0}$ and $v_{0}$, $R_{0}$ the resistance
associated to $V_{0}/I_{0}$ and $S_{I_{0}}$ the shot noise level associated
to $I_{0}$.

To complete our analysis, we will perform MC simulations of the system under
study, so that the validity of the analytical theory can be checked by
comparison. To this end we use an ensemble MC simulator, 3D in momentum
space and self-consistently coupled with a 1D Poisson solver to account for
Coulomb interaction. The carrier dynamics is simulated in the ballistic
active region of the structure and the electron injection from the thermal
reservoirs is modeled according to Fermi statistics. Due to Pauli principle,
the instantaneous occupancy of an incoming electron state with energy $%
\varepsilon $ and impinging at the interface between the ideal thermal
reservoir and the active region fluctuates in time obeying a binomial
distribution\cite{Levitov93} with a probability of success given by $%
f_{FD}(\varepsilon -\mu )$. This statistics is implemented in the MC
simulation of the contact injection by introducing a discretization of
momentum space and using the rejection technique to select the times of
injection at every momentum state.\cite{Gonzalez99b} As limiting cases, when 
$\varepsilon -\mu \ll -1$, $f_{FD}(\varepsilon -\mu )\cong 1$ and the
injection statistics of the corresponding state is uniform in time. By
contrast, when $\varepsilon -\mu \gg 1$, $f_{FD}(\varepsilon -\mu )\ll 1$
and the injection statistics is Poissonian. For the calculations we use the
following parameters: $T=300$ K, $m=0.25\ m_{0}$, $\epsilon =11.7\ \epsilon
_{0}$, with $m_{0}$ the free electron mass and $\epsilon _{0}$ the vacuum
permittivity.

\section{Analytical solution}

\label{solution}

The low frequency solution of the model presented in Sec.~\ref{Model} can be
obtained in fully analytical form. For the sake of conciseness, below we
only present the final expression of the relevant quantities of interest and
refer to the appendix \ref{appendix} for the details of the derivation.

\subsection{$I-V$ characteristics and steady-state profiles}

\label{transport}

Following the results presented in appendix \ref{appendix}, the current
voltage ($I-V$) characteristics of the ballistic conductor presented in Sec.~%
\ref{System} can be calculated as 
\begin{equation}
\overline{I}=\int_{\overline{U}_{m}}^{+\infty }du\left[ f_{c}(u-\xi
_{L})-f_{c}(u-\xi _{0})\right] \text{ ,}  \label{IVav}
\end{equation}
where $f_{c}(\varepsilon )$ is given in Eq.~(\ref{Fbcav}), $\overline{U}_{m}$
is the average value of the maximum potential energy at applied bias $V$,
and $\xi _{L}=V+\mu $ and $\xi _{0}=\mu $ are the electrochemical potentials
at the contacts located at $x=L$ and $x=0$, respectively. The value of $%
\overline{U}_{m}$ can be calculated through the following equation (see
appendix \ref{appendix}) 
\begin{equation}
L=\int_{\overline{U}_{0}}^{\overline{U}_{m}}\frac{dU}{E^{-}\left( U,%
\overline{U}_{m}\right) }-\int_{\overline{U}_{L}}^{\overline{U}_{m}}\frac{dU%
}{E^{+}\left( U,\overline{U}_{m}\right) }\text{ ,}  \label{Umav}
\end{equation}
where 
\begin{eqnarray}
&&E^{-}(U,\overline{U}_{m})=  \nonumber \\
&&\left\{ \int_{\overline{U}_{m}}^{\infty }du\left[ \sqrt{u-U}-\sqrt{u-%
\overline{U}_{m}}\right] \left[ f_{c}(u-\xi _{L})+f_{c}(u-\xi _{0})\right]
\right.  \nonumber \\
&&\left. +\int_{U}^{\overline{U}_{m}}du\sqrt{u-U}2f_{c}(u-\xi _{0})\right\}
^{\frac{1}{2}}2^{\frac{3}{4}}  \label{E-av}
\end{eqnarray}
and 
\begin{eqnarray}
&&E^{+}(U,\overline{U}_{m})=  \nonumber \\
&&-\left\{ \int_{\overline{U}_{m}}^{\infty }du\left[ \sqrt{u-U}-\sqrt{u-%
\overline{U}_{m}}\right] \left[ f_{c}(u-\xi _{L})+f_{c}(u-\xi _{0})\right]
\right.  \nonumber \\
&&\left. +\int_{U}^{\overline{U}_{m}}du\sqrt{u-U}2f_{c}(u-\xi _{L})\right\}
^{\frac{1}{2}}2^{\frac{3}{4}}\text{ .}  \label{E+av}
\end{eqnarray}

\begin{figure}[t]
\centerline{
\mbox{\epsfxsize=8.5cm \epsffile{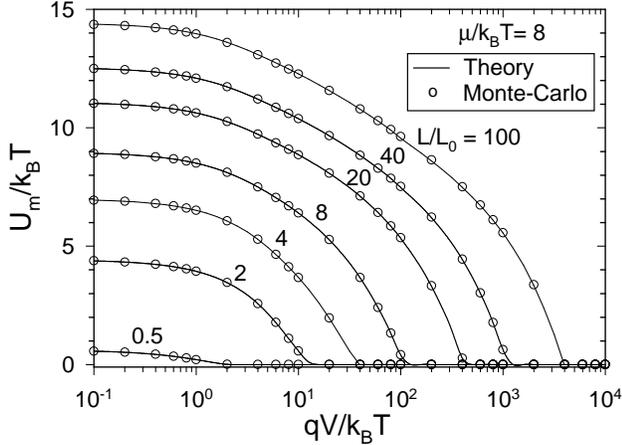}}
}
\caption{Maximum energy potential $\bar{U}_m/k_B T$ as a function of voltage
for $\mu/k_B T=8$ and several values of the sample length $L/L_0$. Solid
line refers to the result of the analytical calculations and open circles to
those of the MC simulations.}
\label{FigUmV}
\end{figure}

Figure ~\ref{FigUmV} reports the maximum potential energy as a function of
the voltage, as calculated from Eq.~(\ref{Umav}) with $\mu =8$ at different
values of the sample length. MC results are also shown. An excellent
agreement is found between analytical and MC calculations. At vanishing
voltages $\overline{U}_{m}$ saturates at higher values the longer is the
normalized length, because of the increasing effect of space charge inside
the ballistic region. At increasing voltages $\overline{U}_{m}$ starts
decreasing till vanishing at a threshold voltage $V_{cr}^{m}$.

\begin{figure}[t]
\centerline{
\mbox{\epsfxsize=8.5cm \epsffile{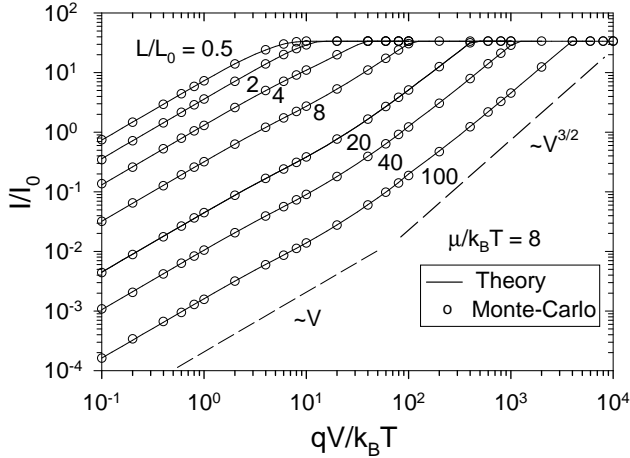}}
}
\caption{I-V characteristics for $\mu/k_B T=8$ and several values of the
sample length $L/L_0$. Solid lines refer to the result of the analytical
calculations and open circles to those of MC simulations. Dashed curves
indicate the typical power law behaviors of the Ohmic ($V$) and
Child-Langmuir ($V^{3/2}$) regimes.}
\label{FigIV}
\end{figure}

Figure ~\ref{FigIV} shows the $I-V$ characteristics as calculated from Eq.~(%
\ref{IVav}) for $\mu =8$ and several sample lengths. The figure also reports
the results of MC simulations. We have found perfect agreement between the
results of the theory and those of MC simulations as expected, since both
results are exact solutions of the same physical model. The main features of
the $I-V$ characteristics consist in the presence of a linear (Ohmic)
behavior at low applied voltages and a current saturation regime at high
voltages. The resistance of the linear regime is given by (in dimensionless
units) 
\begin{equation}
R_{eq}=\frac{1}{f_{c}(\overline{U}_{m}^{eq}-\mu )}\text{ .}  \label{Req}
\end{equation}
In general, this value depends on both the sample length $L$ and the reduced
chemical potential $\mu $. When $\overline{U}_{m}^{eq}\ll |\mu |$ the
resistance takes a value independent of the sample length $%
R_{eq}^{univ}=1/f_{c}(-\mu )$ in agreement with the results of Ref.~[%
\onlinecite{Greiner00}]. Typically this happens for $L\ll 5$ and $\mu \gg 5$%
, which represent the conditions for negligible space charge inside the
ballistic region. In this limit, and when $\mu \gg 1$, $R_{eq}^{univ}$ tends
to the Sharvin contact resistance $R_{S}$. The value of the saturation
current is given by (in dimensionless units): 
\begin{equation}
I_{sat}=\int_{0}^{+\infty }f_{c}(u-\mu )du\text{ .}  \label{Isat}
\end{equation}
Current saturation takes place when all carriers injected from one contact
reach the opposite one, while none of the carriers injected from the other
contact is able to cross the ballistic region. The value of the saturation
current is independent of the sample length since it is only determined by
the emission properties of the contacts. The voltage for the onset of
current saturation, $V_{cr}$, depends on both $L$ and $\mu $, as can be seen
in Fig.~\ref{FigIV}. In the case that $\mu >V_{cr}^{m}$, where $V_{cr}^{m}$
is the bias value at which the maximum of the potential energy disappears
(see Fig.~\ref{FigUmV}), one has $V_{cr}\cong \mu +3$. Otherwise, when $\mu
<V_{cr}^{m}$, one has $V_{cr}=V_{cr}^{m}$, and saturation coincides with the
disappearance of the maximum of the potential energy. Between the linear and
the current saturation regimes the $I-V$ characteristic displays a nonlinear
region, whose properties are determined by the precise values of $L$ and $%
\mu $. The asymptotic behavior of this non-linear region for large values of 
$L$ in the limit when $\overline{U}_{m}\ll V<V_{cr}$ consists of a slightly
sublinear region followed by a superlinear characteristic of Child-Langmuir
type \cite{vanderziel54} $I_{CL}=$ $V^{3/2}/L^{2}$, as has been found in
Ref.~[\onlinecite{Bulashenko01}]. The sublinear region is associated with a
non zero temperature value which smooths out the Fermi distribution. The
superlinear region is associated with space-charge effects driven by the
presence of $U_{m}$.

The steady state profiles can also be calculated in closed analytical form.
The potential energy profile can be obtained in inverse form from the
following relations (see appendix \ref{appendix}): 
\begin{eqnarray}
\int_{U_{0}}^{\overline{U}^{-}(x)}\frac{dU}{E^{-}\left( U,\overline{U}%
_{m}\right) } &=&x\qquad 0<x<\overline{x}_{m}\text{ ,}  \label{Uxav-} \\
\int_{U_{L}}^{\overline{U}^{+}(x)}\frac{dU}{E^{+}\left( U,\overline{U}%
_{m}\right) } &=&x-L\qquad \overline{x}_{m}<x<L\text{ ,}  \label{Uxav+}
\end{eqnarray}
where $\overline{x}_{m}$ is the location of the potential energy maximum
inside the ballistic region. The value of $\overline{x}_{m}$ can be
calculated from either Eq.~(\ref{Uxav-}) or Eq.~(\ref{Uxav+}) by setting the
value of $\overline{U}(x)$ equal to $\overline{U}_{m}$. Once the potential
energy profile is obtained, the electric field profile can be calculated as
(see appendix \ref{appendix}) 
\begin{equation}
\overline{E}(x)=\left\{ 
\begin{array}{c}
E^{-}\left( \overline{U}^{-}(x),\overline{U}_{m}\right) \qquad 0<x<\overline{%
x}_{m} \\ 
E^{+}\left( \overline{U}^{+}(x),\overline{U}_{m}\right) \qquad \overline{x}%
_{m}<x<L
\end{array}
\right. \text{ ,}  \label{Exav}
\end{equation}
where $E^{-}\left( U,U_{m}\right) $ and $E^{+}\left( U,U_{m}\right) $ are
given in Eqs.~(\ref{E-av}) and (\ref{E+av}), respectively. Finally, the
carrier density profile is obtained from (see appendix \ref{appendix}) 
\begin{eqnarray}
&&\overline{n}(x)=  \nonumber \\
&&\left\{ 
\begin{array}{c}
\int_{\overline{U}_{m}}^{+\infty }\frac{du}{\sqrt{2\left( u-\overline{U}%
^{-}(x)\right) }}\left[ f_{c}(u-\xi _{0})+f_{c}(u-\xi _{L})\right] + \\ 
\int_{\overline{U}^{-}(x)}^{\overline{U}_{m}}\frac{du}{\sqrt{2\left( u-%
\overline{U}^{-}(x)\right) }}2f_{c}(u-\xi _{0})\text{;\quad }0<x<\overline{x}%
_{m} \\ 
\int_{\overline{U}_{m}}^{+\infty }\frac{du}{\sqrt{2\left( u-\overline{U}%
^{+}(x)\right) }}\left[ f_{c}(u-\xi _{0})+f_{c}(u-\xi _{L})\right] + \\ 
\int_{\overline{U}^{+}(x)}^{\overline{U}_{m}}\frac{du}{\sqrt{2\left( u-%
\overline{U}^{+}(x)\right) }}2f_{c}(u-\xi _{L})\text{;\quad }\overline{x}%
_{m}<x<L
\end{array}
\right. \text{.}  \label{nxav2}
\end{eqnarray}

\begin{figure}[t]
\centerline{
\mbox{\epsfxsize=8.5cm \epsffile{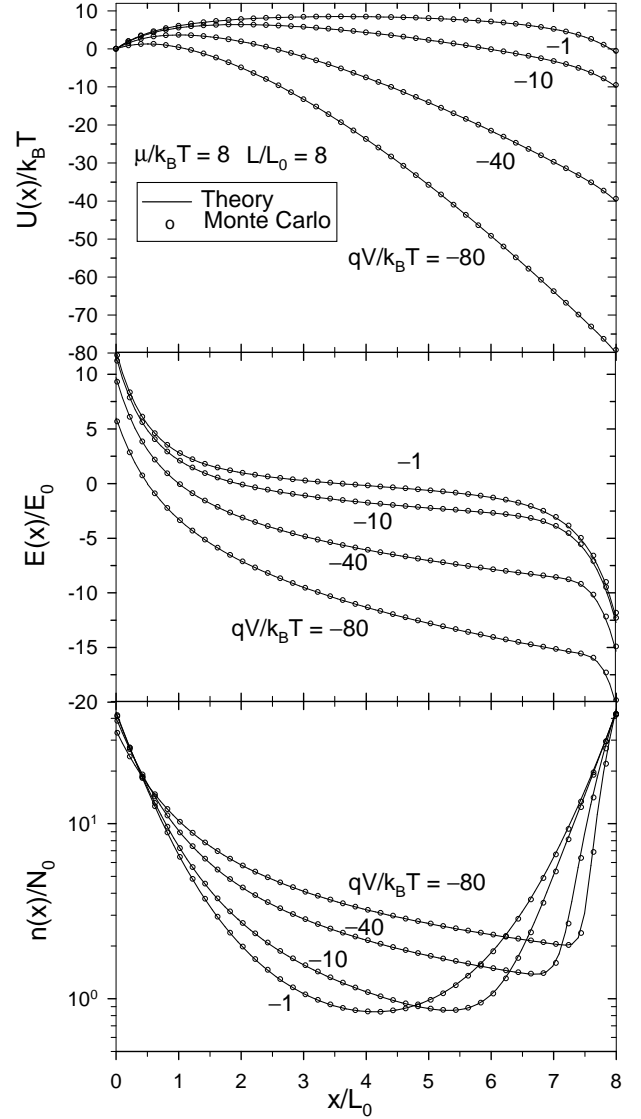}}
}
\caption{Steady state profiles of the potential energy $U(x)/k_BT$, electric
field $E(x)/E_0$ and carrier density $n(x)/N_0$ for $\mu/k_BT=8$ and $L/L_0=8
$, at different values of the applied voltage $qV/k_BT$. Solid lines refer
to the results of analytical calculations, open circles to those of MC
simulations.}
\label{FigProfiles}
\end{figure}

Figure ~\ref{FigProfiles} reports the steady state profiles for $\mu =8$ and 
$L=8$, at several values of the applied bias. For the sake of comparison,
the same figure also reports the profiles obtained from MC simulations.
Again analytical calculations are found to perfectly agree with the
simulations. The figure clearly illustrates the space-charge nature of the
transport and the presence of the potential energy maximum. The value of the
potential energy maximum decreases systematically at increasing bias (as
already indicated in Fig.~\ref{FigUmV}), while its location shifts towards
the left contact. At the same time, the free charge redistributes inside the
structure by shifting the minimum of the carrier density from the center of
the sample towards the right contact.

\subsection{Noise properties}

\label{Noise}

The low frequency noise properties of the system under study are
characterized by the low frequency spectral density of current fluctuations
defined as (in dimensionless units), 
\begin{equation}
S_{I}(0)=\int_{-\infty }^{+\infty }\overline{\delta I(0)\delta I(t)}dt \text{
.}  \label{SIdef}
\end{equation}
When interested in the study of the microscopic correlations the proper
figure of merit is the Fano factor $\gamma $, which is obtained from $%
S_{I}(0)$ as (in the dimensionless units) 
\begin{equation}
\gamma =\frac{S_{I}(0)}{\overline{I}} \text{ .}  \label{Fanodef}
\end{equation}

For conditions far from thermal equilibrium ($V>3$) a value of $\gamma =1$
corresponds to the absence of correlations between different current pulses,
while $\gamma <1$ ($\gamma >1$) corresponds to the presence of negative
(positive) correlations.

As shown in appendix \ref{appendix}, $S_{I}(0)$ can be calculated in closed
analytical form for the model presented in Sec.~\ref{Model}. The final
result can be written in the following compact form (see appendix \ref
{appendix}) 
\begin{eqnarray}
S_{I}(0) &=&\int_{\overline{U}_{L}}^{+\infty }du\left[ \gamma _{L}(u)\right]
^{2}f_{FD}(u-\xi _{L})+  \label{SIav} \\
&&\int_{\overline{U}_{0}}^{+\infty }du\left[ \gamma _{0}(u)\right]
^{2}f_{FD}(u-\xi _{0})\text{ ,}  \nonumber
\end{eqnarray}
where (see appendix \ref{appendix}) 
\begin{eqnarray}
\gamma _{L}(u) &=&\left\{ 
\begin{array}{l}
1+\Omega \widetilde{\gamma }^{>}(u)\qquad \overline{U}_{m}<u<+\infty \\ 
\Omega \widetilde{\gamma }_{L}^{<}(u)\qquad \quad \quad \overline{U}_{L}<u<%
\overline{U}_{m}
\end{array}
\right. \text{,}  \label{gLav} \\
\gamma _{0}(u) &=&\left\{ 
\begin{array}{l}
-1+\Omega \widetilde{\gamma }^{>}(u)\quad \,\,\,\,\overline{U}_{m}<u<+\infty
\\ 
\Omega \widetilde{\gamma }_{0}^{<}(u)\qquad \qquad \,\,\overline{U}_{0}<u<%
\overline{U}_{m}
\end{array}
\right. \text{.}  \label{g0av}
\end{eqnarray}
Here, we have defined 
\begin{equation}
\Omega =-\frac{\left[ \overline{f}_{c}(\overline{U}_{m}-\xi _{L})-\overline{f%
}_{c}(\overline{U}_{m}-\xi _{0})\right] }{\Delta }\text{ ,}  \label{Omega}
\end{equation}
with 
\begin{eqnarray}
\Delta &=&\frac{1}{\overline{E}_{0}}-\frac{1}{\overline{E}_{L}}+\int_{%
\overline{U}_{L}}^{\overline{U}_{m}}du\widetilde{\gamma }_{L}^{<}(u)%
\overline{f}_{FD}(u-\xi _{L})  \nonumber \\
&&+\int_{\overline{U}_{0}}^{\overline{U}_{m}}du\widetilde{\gamma }_{0}^{<}(u)%
\overline{f}_{FD}(u-\xi _{0})+  \nonumber \\
&&+\int_{\overline{U}_{m}}^{+\infty }du\widetilde{\gamma }^{>}(u)\left[ 
\overline{f}_{FD}(u-\xi _{0})+\overline{f}_{FD}(u-\xi _{L})\right] \text{ .}
\label{Deltaav}
\end{eqnarray}
Moreover, 
\begin{eqnarray}
\widetilde{\gamma }^{>}(u) &=&\int_{\overline{U}_{0}}^{\overline{U}_{m}}dU%
\frac{\sqrt{2\left( u-U\right) }-\sqrt{2\left( u-\overline{U}_{m}\right) }}{%
E^{-}\left( U,\overline{U}_{m}\right) ^{3}}  \nonumber \\
&&-\int_{\overline{U}_{L}}^{\overline{U}_{m}}dU\frac{\sqrt{2\left(
u-U\right) }-\sqrt{2\left( u-\overline{U}_{m}\right) }}{E^{+}\left( U,%
\overline{U}_{m}\right) ^{3}}\text{ ,}  \label{g>av} \\
\widetilde{\gamma }_{L}^{<}(u) &=&-2\int_{\overline{U}_{L}}^{u}dU\frac{\sqrt{%
2\left( u-U\right) }}{E^{+}\left( U,\overline{U}_{m}\right) ^{3}}\text{ ,}
\label{g<Lav} \\
\widetilde{\gamma }_{0}^{<}(u) &=&2\int_{\overline{U}_{0}}^{u}dU\frac{\sqrt{%
2\left( u-U\right) }}{E^{-}\left( U,\overline{U}_{m}\right) ^{3}}\text{ .}
\label{g<0av}
\end{eqnarray}
From the previous equations we can evaluate $S_{I}(0)$ and, in turn, the
Fano factor $\gamma $. Note that these analytical expressions are valid in
the whole range of system parameters $L$ and $\mu $, and in the whole range
of applied bias $V$.

\begin{figure}[t]
\centerline{
\mbox{\epsfxsize=8.5cm \epsffile{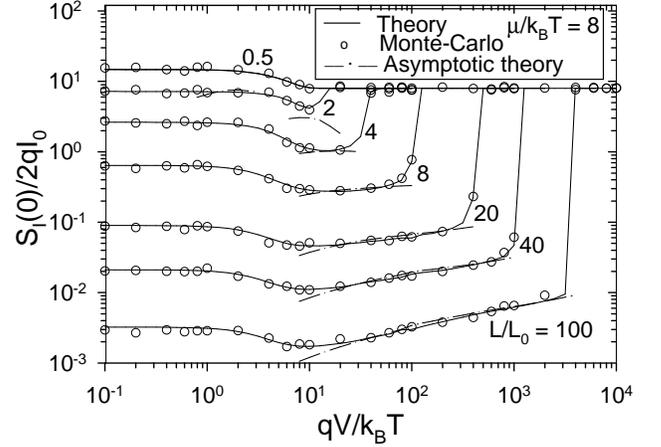}}
}
\caption{Low frequency current spectral density $S_I(0)/2qI_0$ as a function
of the applied voltage $qV/k_BT$ for $\mu/k_BT=8$ and several values of the
sample length $L/L_0$. Solid lines refer to the results of analytical
calculations, open circles to those of MC simulations. Dot-dashed lines are
the results of the asymptotic theory.\protect\cite{Bulashenko01}}
\label{FigSI_V}
\end{figure}

For the purpose of a reliability test, the results obtained from the
analytical formulae are compared with those of MC simulations in Fig. ~\ref
{FigSI_V}. Here, the low frequency current spectral density is reported as a
function of applied bias for $\mu =8$ and several sample lengths. 

As can be
seen in the figure, the agreement between the analytical theory and MC
simulations is excellent, thus proving the reliability of both theory and
simulations. For the sake of completeness Fig.~\ref{FigSI_V} also reports
the results obtained by means of the asymptotic theory developed in Ref.~[%
\onlinecite{Bulashenko01}]. According to this theory, in the limit of $%
\overline{U}_{m}\ll V<V_{cr}$ the low frequency current spectral density can
be approximated by (in dimensionless variables) 
\begin{equation}
S_{I}^{asym}=\beta (\mu -\overline{U}_{m})\frac{I}{V+\overline{U}_{m}}\text{
,}  \label{SIasym}
\end{equation}
with 
\begin{equation}
\beta (\alpha )=9\left( 1-\frac{\pi }{4}\frac{\left[ F_{1/2}(\alpha )\right]
^{2}}{F_{0}(\alpha )F_{1}(\alpha )}\right) \text{ ,}  \label{beta}
\end{equation}
where $F_{j}(\alpha )=1/\Gamma (j+1)\int_{0}^{\infty }dyy^{j}f_{FD}(y-\alpha
)$, with $\Gamma (z)$ being the Gamma function. As seen in Fig.~\ref{FigSI_V}
the asymptotic theory agrees well with the present theory for $\overline{U}%
_{m}\ll V<V_{cr}$. However, it cannot describe either the transition between
low and high bias regions, or the case of small sample lengths in which the
condition $\overline{U}_{m}\ll V<V_{cr}$ can not be satisfied.

The main features of the low frequency spectral density depicted in Fig.~\ref
{FigSI_V} are summarized as follows. At low voltages ($V<3$), where Ohmic
conditions are satisfied, $S_{I}(0)$ is bias independent and takes the value
(in dimensionless variables) 
\begin{equation}
S_{I}^{eq}(0)=\frac{2}{R_{eq}}=2f_{c}(\overline{U}_{m}^{eq}-\mu )\text{ ,}
\label{SIeq}
\end{equation}
in agreement with Nyquist theorem.\cite{nyquist28} At high voltages ($%
V>V_{cr}$), when the sample exhibits current saturation conditions, $%
S_{I}(0) $ is again bias independent and given by (in dimensionless
variables) 
\begin{equation}
S_{I}^{sat}(0)=\int_{0}^{+\infty }duf_{FD}(u-\xi _{0})=f_{c}(-\mu )\text{ .}
\label{SIsat}
\end{equation}
In the region of intermediate values of voltages, $S_{I}(0)$ can present a
monotonic or a non-monotonic behavior (with the presence of a minimum)
determined by the interplay between Coulomb and Pauli correlations, as will
be detailed in the next section.

\section{The scenario of shot-noise in ballistic $n^{+}-i-n^{+}$ structures}

\label{General}

The present theory enables us to investigate separately the relevance of the
two mechanisms responsible for the correlations in the system under study,
namely, the Pauli exclusion principle and the long range Coulomb
interaction, in the whole range of system parameters. Accordingly, we
propose a general scheme which summarizes the whole scenario of the
shot-noise properties exhibited by $n^{+}-i-n^{+}$ ballistic semiconductor
structures. To construct such a scheme, we make use of the Fano factor, $%
\gamma $, which is factorized into the two independent contributions, $%
\gamma _{P}$ and $\gamma _{C}$, related to the Pauli and Coulomb
correlations, respectively. Indeed, according to Eq.~(\ref{Fccorr}), the
fluctuations of the contact distribution function at different energy levels
are uncorrelated, so that the only source of correlations among carriers
injected with different energy is the Coulomb interaction in the active
region. As a consequence, both contributions to the Fano factor are
independent, which implies $\gamma =\gamma _{P}\gamma _{C}$. Thus, the Pauli
contribution $\gamma _{P}$ corresponds to the Fano factor that would be
obtained in the absence of the self-consistent long range Coulomb
interaction. It can be easily evaluated from the noise calculation performed
in the appendix \ref{appendix} by neglecting the self-consistent
contribution. This is equivalent to set up in Eq.~(\ref{SIav}), 
\begin{eqnarray}
\gamma _{L}(u) &=&\left\{ 
\begin{array}{c}
1\qquad \overline{U}_{m}<u<+\infty \\ 
0\qquad \overline{U}_{L}<u<\overline{U}_{m}
\end{array}
\right. \text{ ,}  \label{gLnC} \\
\gamma _{0}(u) &=&\left\{ 
\begin{array}{c}
-1\qquad \overline{U}_{m}<u<+\infty \\ 
0\qquad \overline{U}_{0}<u<\overline{U}_{m}
\end{array}
\right. \text{ .}  \label{g0nC}
\end{eqnarray}
For the current spectral density associated with Pauli correlations only, it
is thus obtained 
\begin{equation}
S_{I}^{P}(0)=f_{c}(\overline{U}_{m}-\xi _{L})+f_{c}(\overline{U}_{m}-\xi
_{0})\text{ ,}  \label{SInC}
\end{equation}
so that the Pauli contribution to the Fano factor is found to be 
\begin{equation}
\gamma _{P}=\frac{S_{I}^{P}(0)}{\overline{I}}=\frac{f_{c}(\overline{U}%
_{m}-\xi _{L})+f_{c}(\overline{U}_{m}-\xi _{0})}{\int_{\overline{U}%
_{m}}^{+\infty }du\left[ f_{c}(u-\xi _{L})-f_{c}(u-\xi _{0})\right] }\text{ ,%
}  \label{FPauli}
\end{equation}
where use is made of Eq.~(\ref{IVav}).

The Coulomb contribution is then evaluated as $\gamma _{C}=\gamma /\gamma
_{P}$. According to these definitions, for $V>3$ values of $\gamma _{P}\neq
1 $ correspond to the presence of Pauli correlations, while values of $%
\gamma _{C}\neq 1$ correspond to the presence of Coulomb correlations.

A detailed analysis of Eq.~(\ref{FPauli}) indicates that the Pauli
contribution is essentially dependent on the difference $\overline{U}%
_{m}-\mu $. Hence, when $\overline{U}_{m}-\mu >0$ it is $\gamma
_{P}\rightarrow 1$, thus indicating the absence of Pauli correlations. In
particular, for nondegenerate injection conditions $\mu <0$, one always has
that $\overline{U}_{m}-\mu >0$, hence indicating the absence of Pauli
correlations, as should be. By contrast, when $\overline{U}_{m}-\mu <0$ it
is $\gamma _{P}<1$, thus indicating the presence of Pauli correlations.
Since $\overline{U}_{m}$ is a decreasing function of the bias, the condition 
$\overline{U}_{m}^{eq}-\mu <0$ implies automatically that the inequality $%
\overline{U}_{m}-\mu <0$ is satisfied for all bias values, and hence the
presence of Pauli correlations for $V>3$. On the contrary, when $\overline{U}%
_{m}^{eq}-\mu >0$ Pauli correlations are absent for low (or intermediate)
bias values and present for bias values sufficiently high to validate the
condition $\overline{U}_{m}-\mu <0$.

Concerning Coulomb correlations, their presence or absence is roughly
determined by a value of the ratio $L/L_{D_{c}}$ higher or lower than unity,
respectively, where $L_{D_{c}}$ is the Debye screening length corresponding
to an homogeneous system with charge density equal to the equilibrium
contact density. Taking into account the effects of degenerancy, $L_{D_{c}}$
is calculated as (in dimensionless variables) 
\begin{equation}
L_{D_{c}}=\left( \frac{dn_{c}^{eq}}{d\mu }\right) ^{-1/2}=\frac{1}{\sqrt{%
2\int_{0}^{+\infty }du\frac{f_{FD}(u-\mu )}{\sqrt{2u}}}}\text{,}  \label{LDc}
\end{equation}
where we used that from Eq.~(\ref{nxav2}) one has 
\begin{equation}
n_{c}^{eq}=\overline{n}^{eq}(0)=2\int_{0}^{+\infty }du\frac{f_{c}(u-\mu )}{%
\sqrt{2u}}\text{ .}  \label{nceq}
\end{equation}

\begin{figure}[t]
\centerline{
\mbox{\epsfxsize=8.5cm \epsffile{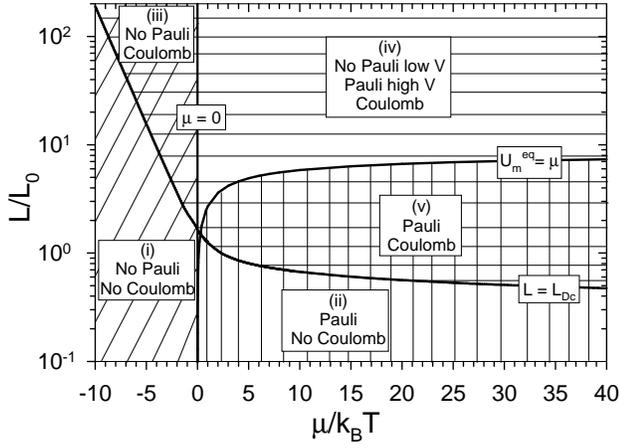}}
}
\caption{Parameter plane in the length /chemical potential space
representing the five different behaviours of the shot-noise properties of a
ballistic conductor according to the relevance of the Pauli and Coulomb
correlations. The three continuous lines define, respectively, the
equalities: $L=L_{D_C}$, $\mu=0$ and $\overline{U}_{m}^{eq}=\mu$.}
\label{FigDiagr}
\end{figure}

On the basis of these considerations, for the scenario of the shot-noise
properties in ballistic conductors we propose the general scheme displayed
in Fig.~\ref{FigDiagr}. In this scheme we identify five different regions in
the plane ($L$, $\mu $), corresponding to five different possibility of
interplay between Coulomb and Pauli correlations. The different regions in
Fig.~\ref{FigDiagr} are determined by the three lines defined by the
equalities $L=L_{D_{c}}$, $\mu =0$ and $\mu= \overline{U}_{m}^{eq}$.

The shot noise behavior in each of the five regions corresponds,
respectively, to: (i) the absence of both Pauli and Coulomb correlations ($%
\mu <0$ and $L/L_{D_{c}}<1$); (ii) the presence of Pauli correlations and
the absence of Coulomb correlations ($\mu >0$, $\overline{U}_{m}^{eq}-\mu <0$
and $L/L_{D_{c}}<1$); (iii) the absence of Pauli correlations and the
presence of Coulomb correlations ($\mu <0$ and $L/L_{D_{c}}>1$); (iv) the
absence (presence) of Pauli correlations for low (high) bias and the
presence of Coulomb correlations ($\mu >0$, $\overline{U}_{m}^{eq}-\mu >0$
and $L/L_{D_{c}}>1$); (v) the presence of both Pauli and Coulomb
correlations ($\mu >0$, $\overline{U}_{m}^{eq}-\mu <0$ and $L/L_{D_{c}}>1$).

\begin{figure}[t]
\centerline{
\mbox{\epsfxsize=8.5cm \epsffile{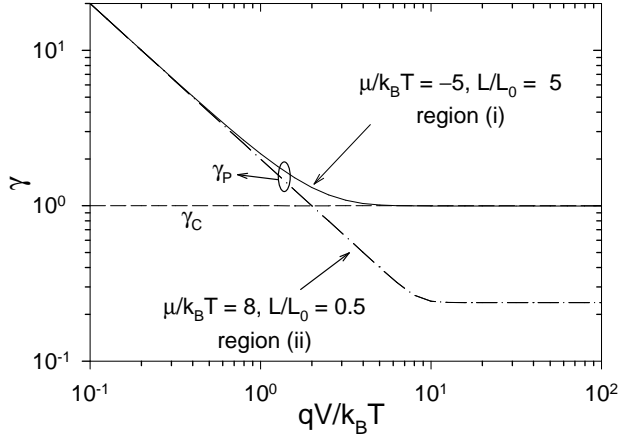}}
}
\caption{Fano factor as a function of the applied voltage $qV/k_BT$ for $%
L/L_0=5$ and $\mu/k_BT=-5$ corresponding to region (i) of Fig.~\ref{FigDiagr}
(solid line) and $L/L_0=0.5$ and $\mu/k_BT=8$ corresponding to region (ii)
of Fig. \ref{FigDiagr} (dot-dashed line). Dashed line represents the Coulomb
contribution to the Fano factor. By definition, in this case the Pauli
contribution to the Fano factor is indistinguishable from the actual Fano
factor.}
\label{FiggL5mu_5}
\end{figure}

\begin{figure}[t]
\centerline{
\mbox{\epsfxsize=8.5cm \epsffile{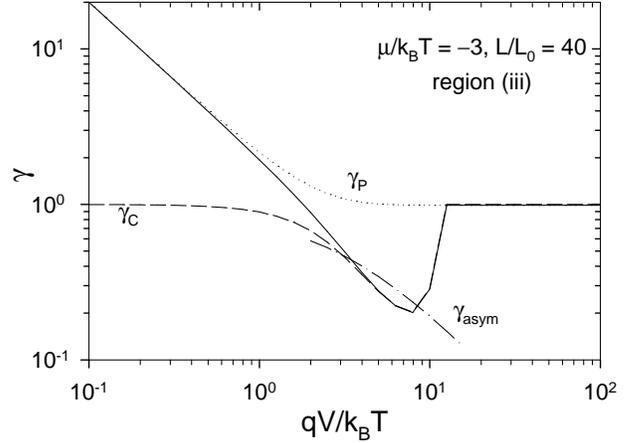}}
}
\caption{Fano factor as a function of the applied voltage $qV/k_BT$ for $%
L/L_0=40$ and $\mu/k_BT=-3$ corresponding to region (iii) of Fig.~\ref
{FigDiagr} (solid line). Dashed and dotted lines represent, respectively,
the Coulomb and Pauli contributions to the Fano factor. Dashed-dotted line
represents the results of the asymptotic theory.}
\label{FiggL40mu_3}
\end{figure}

\begin{figure}[t]
\centerline{
\mbox{\epsfxsize=8.5cm \epsffile{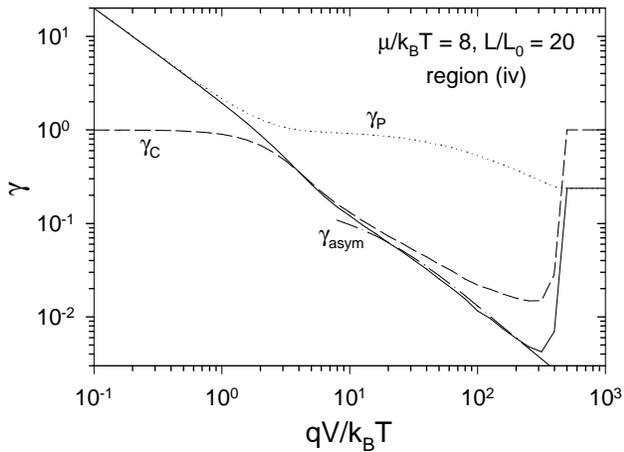}}
}
\caption{Fano factor as a function of the applied voltage $qV/k_BT$ for $%
L/L_0=20$ and $\mu/k_BT=8$ corresponding to region (iv) of Fig.~\ref
{FigDiagr} (solid line). Dashed and dotted lines represent, respectively,
the Coulomb and Pauli contributions to the Fano factor. Dashed-dotted line
represents the results of the asymptotic theory.}
\label{FiggL20mu8}
\end{figure}

\begin{figure}[t]
\centerline{
\mbox{\epsfxsize=8.5cm \epsffile{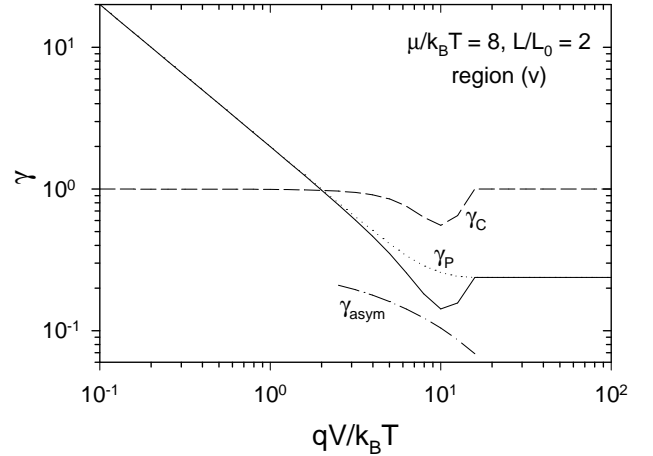}}
}
\caption{Fano factor as a function of the applied voltage $qV/k_BT$ for $%
L/L_0=2$ and $\mu/k_BT=8$ corresponding to region (v) of Fig.~\ref{FigDiagr}
(solid line). Dashed and dotted lines represent, respectively, the Coulomb
and Pauli contributions to the Fano factor. Dashed-dotted line represents
the results of the asymptotic theory.}
\label{FiggL2mu8}
\end{figure}

The reliability of the above scheme has been tested by performing a series
of theoretical calculations for the relevant regions identified above. We
have found that the proposed scheme is essentially valid, except in the
zones close to the lines separating the different regions, where
intermediate behaviors have been observed. Representative examples
concerning the Fano factor, $\gamma $, and the contributions into which it
is decomposed, $\gamma _{P}$ and $\gamma _{C}$, are shown in Figs.~\ref
{FiggL5mu_5} to \ref{FiggL2mu8} for each of the five regions individuated in
the $L$-$\mu $ plane.

Figure~\ref{FiggL5mu_5} displays the Fano factor for $L=5$ and $\mu =-5$ ($%
L/L_{D_{c}}=0.32$) corresponding to region (i) (continuous line) and for $%
L=0.5$ and $\mu =8$ ($L/L_{D_{c}}=0.71$) corresponding to region (ii)
(dot-dashed line). In both cases the presence of space charge in the active
region is nearly negligible, insufficient to originate a potential energy
maximum large enough so as to lead to Coulomb suppression. Therefore, both
regions concern with the absence of Coulomb correlations ($\gamma _{C}=1$).
In region (i) Pauli correlations are absent ($\gamma _{P}=1$ for $V>3$),
since $\mu =-5$ implies that carriers are injected at the contacts with
energies $\varepsilon -\mu >3$, so that the distribution function at the
contacts is well approximated by the nondegenerate Maxwell-Boltzmann
distribution, and the injection statistics is poissonian. On the contrary,
in region (ii) Pauli correlations are responsible for the suppression of
shot noise ($\gamma _{P}<1$, for $V>3$) since degenerate injection
conditions prevail. 

It is worth noting that, in all cases where the long
range Coulomb correlations are absent, as in the present case, the Fano
factor at low voltages decreases inversely with the applied bias according
to the law 
\begin{equation}
\gamma ^{th}=\frac{2}{V}\text{ ,}  \label{FanoTher}
\end{equation}
which corresponds to the thermal noise behavior, while at high voltages ($%
V>V_{cr}$) it becomes constant with a value given by 
\begin{equation}
\gamma ^{sat}=\frac{f_{c}(-\mu )}{\int_{0}^{+\infty }f_{c}(u-\mu )du}\text{ ,%
}  \label{Fanosat}
\end{equation}
which corresponds to current saturation conditions. To obtain Eq. (\ref
{Fanosat}) use is made of Eqs.~(\ref{Isat}) and (\ref{SIsat}). The value of $%
\gamma ^{sat}$ interpolates monotonically between $1$ for $\mu <-3$
(nondegenerate injection statistics) and $2/\mu $ for $\mu >3$ (strongly
degenerate injection statistics). Since the Coulomb correlations vanish
identically at thermal equilibrium and under current saturation conditions,
the two limiting behaviors represented by Eqs.~(\ref{FanoTher}) and (\ref
{Fanosat}) are common to all cases, as we will see below.

Figure \ref{FiggL40mu_3} reports the Fano factor for $L=40$ and $\mu =-3$ ($%
L/L_{D_{c}}=6.9$) corresponding to region (iii) of the general scheme. It
shows the presence of Coulomb correlations ($\gamma _{C}<1$ for $3<V<V_{cr}$%
) and the absence of Pauli correlations ($\gamma _{P}=1$ for $V>3$). Pauli
correlations are absent because $\mu =-3<0$ (see above), while Coulomb
correlations set in due to relevant space charge effects. Due to the small
value of the ratio $L/L_{D_{c}}$ the asymptotic theory\cite{Bulashenko01}
provides only some rough agreement with the exact result in the limited
range of voltages associated with a suppressed behavior. For larger values
of the ratio $L/L_{D_{c}}$ the agreement improves. The behavior displayed by
the system in region (iii) was previously examined in detail in Ref.~[%
\onlinecite{Bulashenko00a}].

Figure \ref{FiggL20mu8} reports the Fano factor for $L=20$ and $\mu =8$ ($%
L/L_{D_{c}}=28.2$) corresponding to the region (iv) of the general scheme.
Here we assist to a low and intermediate voltage range where Coulomb
correlations are dominating and to a high voltage range where Pauli
correlations prevail. Indeed, the Fano factor displays the presence of
Coulomb correlations ($\gamma _{C}<1$ for $3<V<V_{cr}$), the absence of
Pauli correlations in an intermediate bias range ($\gamma _{P}\sim 1$ for $%
3<V\lesssim 10$), and the presence of Pauli correlations for higher bias ($%
\gamma _{P}<1$ for $V\gtrsim 10$). 

Physically, this behavior is due to the
fact that in this region, even if contacts are under degenerate injection
conditions ($\mu >0$), one has $\overline{U}_{m}^{eq}-\mu >0$, so that for
the lowest applied voltages ($3<V\lesssim 10$) it is $\overline{U}_{m}-\mu
>0 $ and the current flows is only due to Poissonian carriers at the tail of
the Fermi distribution function, leading to $\gamma _{P}\sim 1$. As $V$
increases, it is $\overline{U}_{m}-\mu \leq 0$ and sub-Poissonian carriers
near and below the Fermi level increasingly contribute to the current and
low-frequency noise, so that Pauli correlations become manifest in the
noise, thus leading to $\gamma _{P}<1$. Note that in region (iv) one has
generally $L/L_{D_{c}}\gtrsim 10$, thus ensuring that the asymptotic theory 
\cite{Bulashenko01} provides a good approximation to the exact theory
presented here in the range of bias satisfying $U_{m}\ll V<V_{cr}$, as
illustrated in Fig.~\ref{FiggL20mu8}.

Finally, Fig.~\ref{FiggL2mu8} reports the Fano factor for $L=2$ and $\mu =8$
($L/L_{D_{c}}=2.8$) corresponding to region (v) of the general scheme. The
Fano factor displays the presence of both Pauli correlations ($\gamma _{P}<1$
for $V>3$) and Coulomb correlations ($\gamma _{C}<1$ for $3<V<V_{cr}$). In
this region one always has $\overline{U}_{m}-\mu <0$ thus ensuring that
Pauli correlations are present at all applied bias. Therefore, this is the
most interesting region to analyze the interplay between Coulomb and Pauli
correlations. We note that here the exact theory presented in this paper
becomes strictly necessary, since in this region the ratio $L/L_{D_{c}}$
never takes values much higher than unity (one typically obtains $1\lesssim
L/L_{D_{c}}\lesssim 10$), thus reducing significantly the usefulness of the
asymptotic theory,\cite{Bulashenko01} as illustrated in Fig.~\ref{FiggL2mu8}.

\begin{figure}[t]
\centerline{
\mbox{\epsfxsize=8.5cm \epsffile{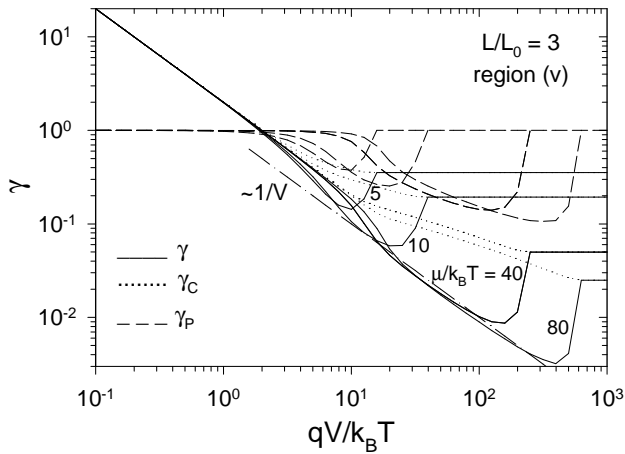}}
}
\caption{Fano factor as a function of the applied voltage $qV/k_BT$ for $%
L/L_0=5$ and several values of $\mu/k_BT$ corresponding to region (v) of
Fig.~\ref{FigDiagr} (solid line). Dashed and dotted lines represent,
respectively, the Coulomb and Pauli contributions to the Fano factor. The
dashed-dotted line gives the $1/V$ slope.}
\label{FiggL5}
\end{figure}

To provide more insight into the interplay between Coulomb and Pauli
correlations in region (v), Fig.~\ref{FiggL5} reports the Fano factor as a
function of bias for $L=3$ and several values of $\mu $ belonging to region
(v). It is observed that by increasing the value of $\mu $ the contribution
of the Pauli correlations decreases faster than that of the Coulomb
correlations. This is understood by noting that for $\mu >3$, $\gamma _{P}$
varies as $2/\mu $, while $\gamma _{C}$, being determined by the ratio $%
L/L_{D_{c}}$, varies as $\sqrt{\mu }$. Physically, this behavior reflects
the increasing amount of sub-Poissonian carriers contributing to the current
as $\mu $ becomes higher. These carriers lead to a more and more pronounced
Pauli suppression while they do not contribute so much to Coulomb
suppression, since due to their sub-Poissonian character the fluctuations of 
$U_{m}$ they induce are much less significant than those originated by the
Poissonian carriers above the Fermi level. For this reason, $\gamma _{C}$
tends to saturate for the highest values of $V<V_{cr}$ while $\gamma _{p}$
continues decreasing. Remarkably enough, in the region (v) for $V>30$ the
total Fano factor $\gamma $ is well described in all cases by a $1/V$
dependence. As seen in the figure, this dependence on voltage is due to the
joint action of both Pauli and Coulomb correlations. It can also be observed
that the onset of Coulomb suppression takes place for higher values of $V$
as $\mu $ increases, since for this suppresion to become significant it is
necessary that the contribution to the current due to carriers injected at
the right contact becomes negligible. This takes place when $V+\overline{U}%
_{m}\cong \mu $, condition that requires a higher value of $V$ as $\mu $
increases, for a given value of $L$.

To illustrate a realistic example where the theoretical results presented
here could be applied, we consider as a significant example the case of a
ballistic GaAs $n^{+}-i-n^{+}$ homodiode at $T=4\ K$ and contact density $%
n_{c}=1.14$ $10^{16}$ cm$^{-3}$. For this structure $k_{B}T/q=0.3$ mV, $\mu
=8$ ($0.24$ meV), $L_{0}\sim 31$ nm and $L_{D_{c}}\sim 0.7$ ($22$ nm), where
we have taken $m=0.066\ m_{0}$, and $\epsilon =12.9\ \epsilon _{0}$. For a
sample length of $500$ nm, $L\sim 16$, and according to Fig.~\ref{FigDiagr}
we are in region (iv) of the general scheme. For a sample length of $70$ nm, 
$L\sim 2.3$ and we are in region (v). For a sample length of $15$ nm, $L\sim
0.5$ and we are in region (ii). To explore regions (i) and (iii) one can
refer to the case of heterodiodes.\cite{Bulashenko99} We conclude that in
principle the different behaviors of the Fano factor predicted by the
present theory can be investigated experimentally within realistic
conditions.

\section{Conclusions}

\label{Conclusions}

We have presented a semiclassical theory of non-equilibrium noise
(shot-noise) properties of $n^{+}-i-n^{+}$ ballistic semiconductor
structures aiming at evaluating the relevance of the long range Coulomb
interaction and of the Pauli exclusion principle. The theory covers the
whole range of system parameters, physically identified in the contact
chemical potential, the sample length, and the applied voltage. Within a
unitary scheme free from any approximation we succeed in investigating the
transition between: nondegenerate and degenerate injection conditions, short
and long sample lengths, and low and high applied bias. Through the
determination of the Fano factor, we have analyzed the relevance of Coulomb
and Pauli correlations. At applied voltages above about $3k_{B}T/q$ both
correlations lead to the suppression of shot noise. We have identified five
different regions in the plane defined by the sample length and the chemical
potential corresponding, respectively, to the following conditions. (i) the
absence of Pauli and Coulomb correlations; (ii) the presence of Pauli
correlations and the absence of Coulomb correlations; (iii) the absence of
Pauli correlations and the presence of Coulomb correlations; (iv) the
absence (presence) of Pauli correlations for low (high) bias and the
presence of Coulomb correlations; and (v) the presence of both Pauli and
Coulomb correlations. Case (i) corresponds to small sample lengths and
non-degenerate conditions when different current pulses are clearly
uncorrelated. Case (ii) occurs for small sample lengths and degenerate
injection conditions so that Pauli correlations are the only being active.
Case (iii) implies a large sample length and non-degenerate injection
conditions so that Coulomb correlations are the only being active. Case (iv)
is associated with large samples and degenerate injection conditions. Here
Coulomb correlations are always present because of the small value of the
Debye screening length. In addition, at low bias only Poissonian carriers at
the tail of the contact energy distribution contribute to the noise so that
Pauli correlations are absent, while at higher bias Pauli principle becomes
active due to the increasing amount of carriers obeying binomial injection
statistics that contribute to noise. Finally, case (v) refers to moderately
long samples and degenerate injection conditions so that both Coulomb and
Pauli correlations are present simultaneously. The results of the theory are
in perfect agreement with analogous MC simulations. Therefore, besides
offering a complete physical picture of the subject, this study provides new
insight into the noise properties of ballistic conductors, and of mesoscopic
systems in general. We believe that the theory here developed constitutes a
powerful tool to design experimental investigations of the nonequilibrium
noise properties of solid-state ballistic conductors.

\section{Acknowledgments}

Partial support from the DURSI of the Generalitat de Catalunya (Spain), the
Consejer\'{\i}a de Educaci\'on y Cultura de la Junta de Castilla y Le\'on
thorugh the project SA057/02, the Ministerio de Ciencia y Tecnolog\'{\i }a
(Spain) through the projects TIC2001-1754, BFM2000-0624 and the Ramon y
Cajal program, and from the Italy-Spain Joint Action of the MIUR Italy (Ref.
IT-109) and MCyT Spain (Ref. HI2000-0138) is gratefully acknowledged.

\appendix

\section{Derivation of the main formulae}

\label{appendix}

In this appendix we derive the main formulae used in the paper by following
a procedure similar to that used in Ref.~[\onlinecite{Bulashenko00a}], here
extended to the case of degenerate injection conditions. To this end, the
first step consists in solving for the distribution function that satifies
the Vlasov-Poisson-Langevin system of equations, Eqs.~(\ref{Vlasov})-(\ref
{Fccorr}). To find the solution we first note that, since transport is
ballistic, the total longitudinal energy of a carrier $u$, defined as 
\begin{equation}
u=\frac{1}{2}v_{x}^{2}+U(x,t)=\varepsilon _{x}+U(x,t)\text{ ,}
\label{utotal}
\end{equation}
remains constant during the flight of the carrier through the structure.
Then, we note that, since transport is ballistic, carriers can reach a given
point inside the structure either directly from the contacts or indirectly
after reflection at the self-consistent potential barrier. Moreover, we can
identify from which contact the carrier comes from. From these
considerations it is easy to convince oneself that, in the low frequency
limit of interest in the present paper, the distribution function solving
the set of Eqs.~(\ref{Vlasov})-(\ref{Fccorr}) is given by 
\begin{eqnarray}
&&F(x,v_{x},t)=  \nonumber \\
&&\left\{ 
\begin{array}{l}
f_{0}(u,t)\quad U_{0}<u<\infty \text{, }v_{x}>0\text{, }0<x<x_{m} \\ 
f_{L}(u,t)\quad U_{m}<u<\infty \text{, }v_{x}<0\text{, }0<x<x_{m} \\ 
f_{0}(u,t)\quad U_{0}<u<U_{m}\text{, }v_{x}<0\text{, }0<x<x_{m} \\ 
f_{0}(u,t)\quad U_{m}<u<\infty \text{, }v_{x}>0\text{, }x_{m}<x<L \\ 
f_{L}(u,t)\quad U_{L}<u<U_{m}\text{, }v_{x}>0\text{, }x_{m}<x<L \\ 
f_{L}(u,t)\quad U_{L}<u<\infty \text{, }v_{x}<0\text{, }x_{m}<x<L
\end{array}
\right. \text{ ,}  \label{Fsol}
\end{eqnarray}
where 
\begin{eqnarray}
f_{0}(u,t) &=&F\left( 0,+\sqrt{2\left( u-U_{0}\right) },t\right) \text{ ,}
\label{fubc0} \\
\text{ }f_{L}(u,t) &=&F\left( L,-\sqrt{2\left( u-U_{L}\right) },t\right) 
\text{ ,}  \label{fubcL}
\end{eqnarray}
with $F\left( 0,v_{x},t\right) $ and $F\left( L,v_{x},t\right) $ being
obtained through the boundary conditions in Eqs.~(\ref{Fbc0}) and (\ref{FbcL}%
). In the previous expression $U_{m}\equiv U_{m}(t)$ and $x_{m}\equiv
x_{m}(t)$ refer to the value and location of the potential energy maximum,
respectively, while $U_{0}$ and $U_{L}$ correspond to the values of the
potential energy at the contacts (note that, according to the boundary
conditions assumed here, these values do not fluctuate).

Since Eq.~(\ref{Fsol}) depends on the value of the maximum potential energy,
to completely determine the distribution function we need to derive the
equation satisfied by the energy maximum. To this purpose, we first obtain
an expression for the carrier density by performing the integral in Eq.~(\ref
{density}) with the help of Eq. (\ref{Fsol}). After some algebra one can
show that the carrier density is given by 
\begin{eqnarray}
&&n(x,t)=  \nonumber \\
&&\left\{ 
\begin{array}{c}
n^{-}\left( U(x,t),U_{m};[f_{0},f_{L}]\right) \,0<x<x_{m} \\ 
n^{+}\left( U(x,t),U_{m};[f_{0},f_{L}]\right) \,\,x_{m}<x<L
\end{array}
\right. \text{,}  \label{nx}
\end{eqnarray}
where 
\begin{eqnarray}
n^{-}(U,U_{m}; &[&f_{0},f_{L}])=  \nonumber \\
&&\int_{U_{m}}^{+\infty }\frac{du}{\sqrt{2(u-U)}}\left[
f_{0}(u,t)+f_{L}(u,t)\right]  \nonumber \\
&&+\int_{U}^{U_{m}}\frac{du}{\sqrt{2(u-U)}}2f_{0}(u,t)  \label{nU-}
\end{eqnarray}
and 
\begin{eqnarray}
n^{+}(U,U_{m}; &[&f_{0},f_{L}])=  \nonumber \\
&&\int_{U_{m}}^{+\infty }\frac{du}{\sqrt{2(u-U)}}\left[
f_{0}(u,t)+f_{L}(u,t)\right]  \nonumber \\
&&+\int_{U}^{U_{m}}\frac{du}{\sqrt{2(u-U)}}2f_{L}(u,t)\text{ .}  \label{nU+}
\end{eqnarray}
In the previous equations the square brackets on the l.h.s. mean a
functional dependence. Now, by multiplying the Poisson equation, Eq.(\ref
{Poisson}), by $\partial U(x,t)/\partial x$ and integrating with respect to $%
x$, it can be shown that the electric field $E(x,t)$ is given through 
\begin{eqnarray}
&&E(x,t)=  \nonumber \\
&&\left\{ 
\begin{array}{c}
E^{-}\left( U(x,t),U_{m};[f_{0},f_{L}]\right) \,0<x<x_{m} \\ 
E^{+}\left( U(x,t),U_{m};[f_{0},f_{L}]\right) \,x_{m}<x<L
\end{array}
\right. \text{,}  \label{Ex}
\end{eqnarray}
with 
\begin{eqnarray}
&&E^{-}(U,U_{m};[f_{0},f_{L}])=  \nonumber \\
&&\left\{ \int_{U_{m}}^{\infty }du\left[ \sqrt{u-U}-\sqrt{u-U_{m}}\right]
\left[ f_{0}(u,t)+f_{L}(u,t)\right] \right.  \nonumber \\
&&\left. +\int_{U}^{U_{m}}du\sqrt{u-U}2f_{0}(u,t)\right\} ^{\frac{1}{2}}2^{%
\frac{3}{4}}  \label{EU-}
\end{eqnarray}
and 
\begin{eqnarray}
&&E^{+}(U,U_{m};[f_{0},f_{L}])=  \nonumber \\
&&-\left\{ \int_{U_{m}}^{\infty }du\left[ \sqrt{u-U}-\sqrt{u-U_{m}}\right]
\left[ f_{0}(u,t)+f_{L}(u,t)\right] \right.  \nonumber \\
&&\left. +\int_{U}^{U_{m}}du\sqrt{u-U}2f_{L}(u,t)\right\} ^{\frac{1}{2}}2^{%
\frac{3}{4}}\text{ .}  \label{EU+}
\end{eqnarray}
Finally, from the definition $\partial U(x,t)/\partial x=E(x,t)$ and its
integration with respect to the space coordinate, we arrive at the following
inverse equations for the potential energy profile 
\begin{equation}
\int_{U_{0}}^{U^{-}(x,t)}\frac{dU}{E^{-}(U,U_{m};[f_{0},f_{L}])}=x\qquad 
\text{ ,}  \label{Ux-}
\end{equation}
valid for $0<x<x_{m}$, and 
\begin{equation}
\int_{U_{L}}^{U^{+}(x,t)}\frac{dU}{E^{+}(U,U_{m};[f_{0},f_{L}])}=x-L\text{ ,}
\label{Ux+}
\end{equation}
valid for $x_{m}<x<L$.

From the previous expressions the location of the maximum energy potential
can be obtained by evaluating either Eq.~(\ref{Ux-}) or Eq.~(\ref{Ux+}) at $%
x=x_{m}$, 
\begin{eqnarray}
\int_{U_{0}}^{U_{m}}\frac{dU}{E^{-}(U,U_{m};[f_{0},f_{L}])} &=&x_{m}\text{ ,}
\label{xm1} \\
\int_{U_{L}}^{U_{m}}\frac{dU}{E^{+}(U,U_{m};[f_{0},f_{L}])} &=&x_{m}-L\text{
.}  \label{xm2}
\end{eqnarray}
By eliminating $x_{m}$ from the two resulting equations we derive a closed
equation for the value of the potential energy maximum in the form 
\begin{eqnarray}
L &=&\int_{U_{0}}^{U_{m}}\frac{dU}{E^{-}(U,U_{m};[f_{0},f_{L}])}-  \nonumber
\\
&&\int_{U_{L}}^{U_{m}}\frac{dU}{E^{+}(U,U_{m};[f_{0},f_{L}])}\text{ .}
\label{Um}
\end{eqnarray}
Equation (\ref{Um}) constitutes a closed equation to determine the value of
the potential energy maximum, $U_{m}$. Notice that it depends solely on the
boundary conditions for the distribution function and the sample length.

Once the value of the potential energy maximum is known, one can determine
its location through either Eq.~(\ref{xm1}) or Eq. (\ref{xm2}). Then, from
Eq.~(\ref{Fsol}) one obtains the explicit expression of the distribution
function. In a similar way the explicit spatial dependence of the potential
energy can be determined by substituting the value of $U_{m}$ in Eqs.~(\ref
{Ux-}) and (\ref{Ux+}), and that of the electric field and carrier density
by substituting $U_{m}$ in Eq.~(\ref{Ex}) and Eq.~(\ref{nx}), respectively.
In this way, a complete analytical solution of the model presented in Sec.%
\ref{Model} is obtained.

In particular, the electrical current, defined as 
\begin{equation}
I(t)=-\int_{-\infty }^{+\infty }dv_{x}v_{x}F(x,v_{x},t)\text{ ,}
\label{Itdef}
\end{equation}
can be shown to be given by 
\begin{equation}
I(t)=\int_{U_{m}}^{+\infty }du\left[ f_{L}(u,t)-f_{0}(u,t)\right] \text{ .}
\label{IVt}
\end{equation}
Now, we are in the position to derive the corresponding expressions for the
transport and noise properties.

\subsection{Transport properties}

The average steady-state transport properties can be computed directly from
the equations presented above by simply substituting into them 
\begin{eqnarray}
U_{m} &\rightarrow &\overline{U}_{m}\text{ ,}  \nonumber \\
x_{m} &\rightarrow &\overline{x}_{m}\text{ ,}  \label{canvis} \\
f_{0}(u,t) &\rightarrow &\bar{f}_{0}(u)=f_{c}(u-\xi _{0})\text{ ,}  \nonumber
\\
f_{L}(u,t) &\rightarrow &\bar{f}_{L}(u)=f_{c}(u-\xi _{L})\text{ ,}  \nonumber
\end{eqnarray}
with $f_{c}(\varepsilon )$ given through Eq.~(\ref{Fbcav}). In this way we
arrive at the equations used in Sec.~\ref{transport}

\subsection{Low frequency current noise properties}

The fluctuating properties of any of the quantities of interest can be
evaluated directly from the equations derived above by just performing the
corresponding perturbation around the steady state, and by taking into
account that the only source of fluctuations in the system is located at the
contacts, and represented by the fluctuating term in Eqs.~(\ref{Fbcav}). In
this paper we are interested on the low frequency current fluctuations. To
compute them we perturb Eq.~(\ref{IVt}) around the steady state thus
obtaining 
\begin{eqnarray}
\delta I(t) &=&\int_{\overline{U}_{m}}^{+\infty }du\left[ \delta
f_{L}(u,t)-\delta f_{0}(u,t)\right] -  \nonumber \\
&&\left[ \overline{f}_{L}(\overline{U}_{m})-\overline{f}_{0}(\overline{U}%
_{m})\right] \delta U_{m}(t)\text{ .}  \label{dIt}
\end{eqnarray}
where $\delta U_{m}(t)$ represents the fluctuations of the potential energy
maximum. In the above expression we distinguish two contributions to the
current fluctuation, one coming directly from the contacts and the other
coming indirectly through the self-consistent potential fluctuations. To
express the dependence of the second contribution on the noise sources, we
perturb Eq.~(\ref{Um}). In performing such a perturbation it is convenient
to shift all the energy integration variables by an amount equal to $U_{m}$.
After some algebra one then arrives at the following expression for the
fluctuations of the maximum potential energy: 
\begin{eqnarray}
\delta U_{m}(t) &=&\frac{1}{\Delta }\int_{U_{L}}^{+\infty }du\widetilde{%
\gamma }_{L}(u)\delta f_{L}(u,t)+  \nonumber \\
&&\frac{1}{\Delta }\int_{U_{0}}^{+\infty }du\widetilde{\gamma }_{0}(u)\delta
f_{0}(u,t)\text{ ,}  \label{dUm}
\end{eqnarray}
with 
\begin{eqnarray}
\Delta &=&\frac{1}{\overline{E}_{L}}-\frac{1}{\overline{E}_{0}}%
+\int_{U_{L}}^{+\infty }du\widetilde{\gamma }_{L}(u)\overline{f}_{FD}(u-\xi
_{L})  \nonumber \\
&&+\int_{U_{0}}^{+\infty }du\widetilde{\gamma }_{0}(u)\overline{f}%
_{FD}(u-\xi _{0})\text{ .}  \label{Delta}
\end{eqnarray}
where we have used that $\overline{f}_{0}^{\prime }(u)=-\overline{f}%
_{FD}(u-\xi _{0})$ and $\overline{f}_{L}^{\prime }(u)=-\overline{f}%
_{FD}(u-\xi _{L})$. Here, we have defined 
\begin{eqnarray}
\widetilde{\gamma }_{L}(u) &=&\widetilde{\gamma }^{>}(u)\theta \left( u-%
\overline{U}_{m}\right) +\widetilde{\gamma }_{L}^{<}(u)\theta \left( 
\overline{U}_{m}-u\right) \text{ ,}  \label{gtild_Lt} \\
\widetilde{\gamma }_{0}(u) &=&\widetilde{\gamma }^{>}(u)\theta \left( u-%
\overline{U}_{m}\right) +\widetilde{\gamma }_{0}^{<}(u)\theta \left( 
\overline{U}_{m}-u\right) \text{ ,}  \label{gtild_0t}
\end{eqnarray}
with 
\begin{eqnarray}
&&\widetilde{\gamma }^{>}(u)=\int_{U_{0}}^{\overline{U}_{m}}\frac{\left[ 
\sqrt{2\left( u-U\right) }-\sqrt{2\left( u-\overline{U}_{m}\right) }\right]
dU}{E\left( U,\overline{U}_{m};[\overline{f}_{0},\overline{f}_{L}]\right)
^{3}}  \nonumber \\
&&-\text{ }\int_{U_{L}}^{\overline{U}_{m}}\frac{\left[ \sqrt{2\left(
u-U\right) }-\sqrt{2\left( u-\overline{U}_{m}\right) }\right] dU}{E\left( U,%
\overline{U}_{m};[\overline{f}_{0},\overline{f}_{L}]\right) ^{3}}\text{,}
\label{g>t}
\end{eqnarray}
\begin{eqnarray}
\widetilde{\gamma }_{L}^{<}(u) &=&\int_{U_{L}}^{u}\frac{-2\sqrt{2\left(
u-U\right) }dU}{E^{+}\left( U,\overline{U}_{m};[\overline{f}_{0},\overline{f}%
_{L}]\right) ^{3}}\text{ ,}  \label{gtildL} \\
\widetilde{\gamma }_{0}^{<}(u) &=&\int_{U_{0}}^{u}\frac{2\sqrt{2\left(
u-U\right) }dU}{E^{-}\left( U,\overline{U}_{m};[\overline{f}_{0},\overline{f}%
_{L}]\right) ^{3}}\text{ ,}  \label{gtild0}
\end{eqnarray}
By substituting Eq.~(\ref{dUm}) in Eq.~(\ref{dIt}) for the current
fluctuations we finally obtain 
\begin{eqnarray}
\delta I(t) &=&\int_{U_{L}}^{+\infty }du\gamma _{L}(u)\delta f_{L}(u,t)
\label{dItfin} \\
&&+\int_{U_{0}}^{+\infty }du\gamma _{0}(u)\delta f_{0}(u,t)\text{ ,}
\end{eqnarray}
with 
\begin{eqnarray}
\gamma _{L}(u) &=&\theta (u-\overline{U}_{m})+\Omega \widetilde{\gamma }%
_{L}(u)\text{ ,}  \label{gLt} \\
\gamma _{0}(u) &=&-\theta (u-\overline{U}_{m})+\Omega \widetilde{\gamma }%
_{0}(u)\text{ ,}  \label{g0t}
\end{eqnarray}
where 
\begin{equation}
\Omega =-\frac{\left[ \overline{f}_{c}(\overline{U}_{m}-\xi _{L})-\overline{f%
}_{c}(\overline{U}_{m}-\xi _{0})\right] }{\Delta }\text{ .}  \label{Omegat}
\end{equation}
Now we are in the position to compute the low frequency current spectral
density, defined as 
\begin{equation}
S_{I}(0)=\int_{-\infty }^{+\infty }\overline{\delta I(0)\delta I(t)}dt\text{
.}  \label{SIdeft}
\end{equation}
By substituting Eq.~(\ref{dItfin}) into Eq.~(\ref{SIdeft}) we arrive at 
\begin{eqnarray}
S_{I}(0) &=&\int_{\overline{U}_{L}}^{+\infty }du\gamma
_{L}(u)^{2}f_{FD}(u-\xi _{L})+  \nonumber \\
&&\int_{\overline{U}_{0}}^{+\infty }du\gamma _{0}(u)^{2}f_{FD}(u-\xi _{0})%
\text{ ,}  \label{SIt}
\end{eqnarray}
where we have used that from Eq.~(\ref{Fccorr}) one has 
\begin{eqnarray}
&&2\int_{-\infty }^{+\infty }\overline{\delta f_{a}(u,t)\delta f_{a^{\prime
}}(u^{^{\prime }},t^{\prime })}dt^{\prime }=  \nonumber \\
&&f_{FD}\left( u-\xi _{_{a}}\right) \delta _{a,a^{\prime }}\delta
(u-u^{\prime })\delta (t-t^{\prime })\text{ .}  \label{corr_u}
\end{eqnarray}
From Eqs.~(\ref{g>t})-(\ref{SIt}) it is straightforward to arrive at the
equations used in Sec.~\ref{Noise}.

\end{document}